\documentclass{iopart}
\usepackage{graphicx}  
\usepackage{latexsym}
\usepackage{amssymb}

\jl{6}        
\eqnobysec    

\def\beq{\begin{equation}} \def\eeq{\end{equation}}

\def\rmd{{\rm d}} \def\rmD{{\rm D}}

\begin{document}

\title[Frenet-Serret formalism for null world lines]
{Frenet-Serret formalism for null world lines}

\author{
Donato Bini$^* {}^\S {}^\dag$, Andrea Geralico${}^\S$ and Robert T. Jantzen${}^\P {}^\S$
}
\address{
  ${}^*$\
Istituto per le Applicazioni del Calcolo ``M. Picone'', CNR I-00161 Rome, Italy
}

\address{
  ${}^\S$\
  ICRA,
  University of Rome, I-00185 Rome, Italy
}

\address{
  ${}^\dag$\
INFN Sezione di Firenze, I--50019 Sesto Fiorentino (FI), Italy
}

\address{
  ${}^{\P}$\
Department of Mathematical Sciences, Villanova University, Villanova, PA 19085 USA.
}

\begin{abstract}
The Frenet-Serret curve analysis is extended from nonnull to null trajectories in a generic spacetime using the Newman-Penrose formalism, recovering old results which are not well known and clarifying the associated Fermi-Walker transport which has been left largely unexplored in the literature.
This machinery is then used to discuss null circular orbits in stationary axisymmetric spacetimes using the Kerr spacetime as a concrete example, and to integrate the equations of parallel transport along null geodesics in any spacetime.
\end{abstract}

\pacno{04.20.Cv}

\section{Introduction}

The Frenet-Serret formalism championed by Synge for timelike world lines \cite{synge1956,synge1960} is a powerful tool for studying the motion of nonzero rest mass test particles in a given gravitational field. This approach depends only on the geometrical (i.e., intrinsic) properties of the timelike world line of the test particle, independent of any particular coordinate system or observer. 
In particular it has proven to be very useful in understanding the properties of timelike circular orbits in stationary axisymmetric spacetimes
and of Fermi-Walker transport along them, and helps visualize the geometry of this family of orbits.

Castagnino \cite{castagnino1965} (under the influence of Cattaneo and Ferrarese in Rome) extended the Frenet-Serret approach to null trajectories starting from the original work of Cartan \cite{cartan} for null curves in a complex 3-dimensional manifold, including the closely associated Fermi-Walker transport. His work was soon followed up by Bonnor \cite{bonnor1969}, who studied null curves in Minkowski spacetime using this machinery, and Synge \cite{synge1972}, who returned the discussion to a general spacetime and related the Frenet-Serret frame to the successive derivatives of the tangent vector to the null curve, and then returned to the topic much later to interpret the natural parametrization of nongeodesic null world lines in terms of observer 3-space quantities \cite{synge1985}. Urbantke \cite{urbantke1989} later generalized Bonnor's work to conformally flat spacetimes.
Recently the question of Fermi-Walker transport along null curves has been revisited for vectors orthogonal to the tangent vector of the curve \cite{samraj2000}. In the present article, these topics are reformulated and clarified using the Newman-Penrose approach \cite{NP} built specifically to deal with null phenomena, and then applied to null Killing trajectories in stationary axisymmetric spacetimes. Finally parallel transport along null geodesics in any spacetime is examined using these tools. 
We follow the conventions of Chandrasekhar \cite{ChandraBook} with the spacetime signature $+$\,$-$\,$-$\,$-$.

For any frame $\{E_a\}$ ($a=1,2,3,4$) with dual frame $\{W^a\}$
defined along a trajectory in spacetime which is characterized by a constant matrix of inner products, i.e., constant frame components $g_{a b} =E_a \cdot E_b$ of the metric, the connection component matrix-valued 1-form evaluated along the tangent $X^a =\rmd x^a/\rmd\lambda$ to the trajectory (with respect to some parameter $\lambda$)
\beq
  \rmD E_a/ d\lambda = E_b C^b{}_a\ ,\quad 
  C^b{}_a =\Gamma^b{}_{a c} X^c
\eeq
must be antisymmetric when index-lowered using the metric
\begin{eqnarray}
 0 &=& \rmD/\rmd\lambda \, ( E_a \cdot E_b)
   = (\rmD E_a/\rmd\lambda) \cdot E_b + E_a \cdot ( \rmD E_b/\rmd\lambda )
\nonumber\\
   &=& C^c{}_a g_{c b} +  g_{a c}C^c{}_b
   = C_{b a} + C_{a b}\ ,
\end{eqnarray}
i.e., defines a 2-form or equivalently a 2-vector
\beq
  C^\flat = {\textstyle\frac12} C_{a b} W^a\wedge W^b\ ,\
  C^\sharp = {\textstyle\frac12} C^{a b} E_a\wedge E_b\ ,
\eeq
where  $T^\flat$ and $T^\sharp$ indicate the totally covariant and totally contravariant forms of a tensor $T$ obtained by index shifting with the metric.
This 2-form is therefore the generator of a Lorentz transformation of the tangent space, although its matrix representation depends on the choice of metric component matrix $(g_{a b})$.

When the first vector $E_1$ is proportional to the tangent to the trajectory and the tangent is timelike, one can introduce successively 3 spacelike vectors defining the curvature properties of the trajectory which form an orthonormal frame 
for which the Greek index notation $\alpha=0,1,2,3$ is more appropriate to distinguish the timelike frame vector from the rest and
$(g_{\alpha\beta}) = (\eta_{\alpha\beta}) = {\rm diag}(1,-1,-1,-1)$ 
leading to the usual Frenet-Serret scenario employed by Synge \cite{synge1956,synge1960} for various relativistic calculations and modernized by Honig, Schucking and Vishweshwara \cite{honschvis}, while if the tangent is spacelike, the orthonormal frame which results has a number of possibilities for where the plus sign in the signature falls, but the situation is analogous unless one runs into a null vector in the process of defining the remaining vectors in the frame (the derivative of a spacelike unit vector must be orthogonal to that vector, but could
be spacelike, timelike or null, 
unlike the timelike case, where the derivative of a timelike unit vector must be spacelike). When the frame is orthonormal, one can decompose the 2-form $C^\flat$ into its electric and magnetic parts corresponding to the rates of change of the Lorentz boost and rotation of the frame relative to a parallel propagated frame along the trajectory. The corresponding (mixed) tensor $C$ will generate this family of Lorentz transformations.

When $E_1$ is instead a null vector, one can play a similar game but is then led to a 
a quasi-orthogonal tetrad \cite{wong}, or in more modern language, a real null frame (or tetrad for dimension 4) \cite{exactsols}, which is facilitated by the Newman-Penrose formalism based on the closely associated complex null frame. Such a real or complex null frame has a naturally associated orthonormal frame (related by a simple constant linear transformation) 
which can only undergo Lorentz transformations along the curve relative to a parallel transported such frame, the mixed tensor $C$ will be the generator of a family of Lorentz transformations (whose frame component matrices are related by the same constant linear transformation to the usual component matrices with respect to the orthonormal frame).

In the present article we construct a Frenet-Serret-like frame adapted to a null world line with tangent along $l$ and use the Newman Penrose formalism to 
derive a set of equations for the frame vectors in a form similar to Eqs.~(\ref{FSeqs}), introducing new intrinsic quantities which play the same role as the more familiar curvature and torsions, leading to a real null frame. Re-examining the study of null world lines in Minkowski spacetime undertaken by Bonnor \cite{bonnor1969} gives some feeling for what these quantities represent. 
We then apply the results to study the motion of massless test particles along special orbits in the Kerr spacetime, namely along the two repeated principal null directions of the spacetime and along null circular orbits.

The new feature of null curves that few of us have intuition about is that while 2-dimensional circles (in $E_2$) and pseudocircles (in $M_2$) of best fit in the osculating 2-plane are the obvious candidates for comparing the curvature of spacelike and timelike curves respectively (constant curvature curves in the tangent space which are orbits of Killing vector fields associated with the Poincare symmetry of that space), with null curves one is forced into at least 3 dimensions (in $M_3$) where there are two classes of comparison null curves of constant curvature in the tangent space: null helices and null cubics, with nonzero and zero torsions respectively. In $M_2$, null curves must be straight since the null cone is 1-dimensional, but in $M_3$ a null vector has ``room to move" on the null cone and so can change direction, corresponding to acceleration, and leads to a 3-dimensional osculating hyperplane describing the local turning motion in $M_4$.
Although there is a long history of studying the geometry of null curves in Lorentzian spacetimes, little of this seems to have entered the more mainstream knowledge base of general relativity. Coupled with the fact that the null rotations which their tangent vectors may undergo relative to parallel transport (in addition to the usual familiar rotations) are also less known intuitively, the topic has remained sidelined.

\section{The Frenet-Serret frame for a timelike world line reviewed}

Given a single timelike test particle world line with some parametrization
$x^\alpha =x^\alpha(\lambda)$ and corresponding tangent $\rmd x^\alpha/\rmd\lambda$ with squared length 
$(\rmd\tau/\rmd\lambda)^2=g_{\alpha\beta} (\rmd x^\alpha/\rmd\lambda) (\rmd x^\beta/\rmd\lambda) 
=(\rmd x/\rmd\lambda) \cdot (\rmd x/\rmd\lambda)$, one may reparametrize it by the proper time along the world line 
$ \rmd\tau/\rmd\lambda =(\rmd x/\rmd\lambda \cdot \rmd x/\rmd\lambda)^{1/2}$. This parametrization is defined modulo an initial value, i.e., up to an additive constant.
The spacetime Frenet-Serret frame  
with unit tangent (4-velocity) $E_0=U = \rmd x/\rmd\tau $, is then described by the following system of evolution equations \cite{iyer-vish}
\begin{eqnarray}
\fl
\label{FSeqs}
\frac{\rmD E_0}{\rmd\tau }&=\kappa E_1\ ,\ 
\frac{\rmD E_1}{\rmd\tau }=\kappa E_0+\tau_1 E_2\ ,\ 
\frac{\rmD E_2}{\rmd\tau }
=-\tau_1 E_1+\tau_2 E_3\ ,\ 
\frac{\rmD E_3}{\rmd\tau }=-\tau_2 E_2\ .
\end{eqnarray}
This corresponds to the matrices $(g_{\alpha\beta})={\rm diag}(1,-1,-1,-1)$ and
\beq\fl
   (C^\alpha{}_\beta) =\pmatrix{0&\kappa&0&0\cr \kappa&0&-\tau_1&0\cr 0&\tau_1&0&-\tau_2\cr 0&0&\tau_2&0\cr}\ ,\quad
   (C^{\alpha\beta}) =\pmatrix{0&-\kappa&0&0\cr \kappa&0&\tau_1&0\cr 0&-\tau_1&0&\tau_2\cr 0&0&-\tau_2&0\cr}\ .
\eeq
Making the obvious decomposition $C = C_\kappa + C_\tau$ into curvature and torsion parts (the electric and magnetic parts of the corresponding covariant or contravariant tensors),
the associated contravariant tensor is the sum of two simple bivectors
\beq
  C_\kappa^\sharp = \kappa E_1\wedge E_0 = a\wedge U\ ,\quad
  C_\tau^\sharp = (\tau_1 E_1 - \tau_2 E_3)\wedge E_2\ ,
\eeq
the first of a timelike and spacelike vector generating a family of boosts in the plane of the two and the second of two spacelike vectors generating a rotation in their plane.

The absolute value of the curvature $\kappa$ is the magnitude of
the acceleration $a = \rmD U/\rmd\tau =\kappa E_1$, while
the first
and second torsions $\tau_1$ and $\tau_2$ are the components of the Frenet-Serret angular velocity vector obtained by taking the spatial dual of the bivector $C_\tau^\sharp$
\beq
\label{omegaFS}
\omega_{\rm (FS)}=\tau_1 E_3 + \tau_2 E_1\ , \qquad ||\omega_{\rm (FS)}||=[\tau_1^2 + \tau_2^2]^{1/2}\ .
\eeq
The latter determines the angular velocity of rotation of the spatial Frenet-Serret frame  $\{E_1,E_2,E_3\}$ and the acceleration determines the rate of change of the boost of this frame relative to a parallel transported frame along $U$ according to 
\beq
\frac{\rmD E_i}{\rmd\tau } =\omega_{\rm (FS)}\times E_i+\kappa E_0\,\delta^1_i\ ,\ i=1,2,3,
\eeq
where $\times$ is the usual cross product in the subspace of the spatial frame.
Along a Killing trajectory all Frenet-Serret curvature and torsions are constant and 
the problem of evaluating such a frame has been explicitly solved by Iyer and Vishveshwara \cite{iyer-vish}. One can then evaluate the parallel transport along the world line using the matrix exponential \cite{holotaubnut}.

The invariants of the matrix $(C^\alpha{}_\beta)$  are given by
\begin{eqnarray}\fl
I_1=\frac{1}{2}C_{\alpha\beta}C^{\alpha\beta}=
-\kappa^2+\tau_1^2+\tau_2^2\ , \qquad
I_2=\frac{1}{2}C_{\alpha\beta}{}^*C^{\alpha\beta}=-2\kappa\tau_2\ ,
\end{eqnarray} 
where
\beq
({}^*C^{\alpha\beta}) = (\frac12
\eta^{\alpha\beta\gamma\delta}C_{\gamma\delta})
=
\pmatrix{
0&-\tau_2&0&-\tau_1\cr
\tau_2&0&0&0\cr
0&0&0&-\kappa\cr
\tau_1&0&\kappa&0\cr 
}
\ .
\eeq 
Here invariance means that under Lorentz transformations of the tangent space which map the given Frenet-Serret orthonormal frame to any other such orthonormal frame, this component matrix transforms under a Lie algebra adjoint transformation but these invariants do not change and so can be used to classify the orbits of the Lie algebra into equivalence classes.
The vanishing or nonvanishing of these two Lorentz invariants classifies the type of Lorentz transformation they generate \cite{synge1956,holotaubnut}. For nongeodesics ($\kappa\ne0$), the vanishing or nonvanishing of the second invariant $I_2$ is equivalent to the vanishing or nonvanishing of $\tau_2$, in which case the generators are called singular/semi-singular or general (nonsingular) respectively. General Lorentz generators generate a Lorentz 4-screw: a rotation in one spacelike 2-plane and a boost in an orthogonal timelike 2-plane, while semi-singular generators correspond to either a pure rotation or boost alone in a spacelike or timelike 2-plane respectively and singular generators correspond to a null rotation in a timelike 3-plane.

The reciprocal of the curvature $\kappa$ defines a radius of curvature $\mathcal{R}=1/\kappa$. For a spacelike curve with a spacelike first normal, this is associated with an osculating circle of radius $\mathcal{R}$ in the tangent space whose center is a distance $\mathcal{R}$ along the first normal $E_1$. For a timelike curve being considered here, this is instead associated with an osculating pseudocircle (hyperbola) in the tangent space whose center is a distance $\mathcal{R}$ along $-E_1$. For a null curve, one is forced into at least a 3-dimensional subspace of the tangent space (the osculating hyperplane) where our intuition must be extended by studying the simplest constant curvature null curves in $M_3$.

The matrix $C^\alpha{}_\beta$ (correspondingly the tensor $C=C^\alpha{}_\beta E_\alpha \otimes W^\beta$)
determines the parallel transport of vectors along the world line when expressed in terms of frame components $X=X^\alpha E_\alpha$
\beq
  \rmD X^\alpha/\rmd\tau  = \rmd X^\alpha/\rmd\tau  + C^\alpha{}_\beta X^\beta = 0\ .
\eeq
It consists of the sum of two parts
$C = C_\kappa + C_\tau$,
the first associated with the curvature, which is responsible for boosting $E_0$ away from a parallel transported direction, and one associated with the two torsions, which are responsible for the rotation of the plane perpendicular to the plane of the motion determined by $E_0\wedge E_1$. Fermi-Walker transport along the world line is obtained from parallel transport by removing the boost piece of the parallel transport to allow the tangent vector to remain tangent to the world line and allow the local rest space of the timelike observer to stay orthogonal to the world line under the new transport. In other words the new transport should only
boost vectors in the plane of the tangent and its derivative (velocity acceleration plane) relative to parallel transport to realign the timelike direction with the tangent to the curve and nothing more. Fermi-Walker transport of a vector is therefore defined by 
\beq
  \rmD_{\rm(fw)} X^\alpha/\rmd\tau  = \rmD X^\alpha/\rmd\tau  - C_\kappa{}^\alpha{}_\beta X^\beta = 0\ ,
\eeq
or
\beq
\rmD X^\alpha/\rmd\tau  = (a^\alpha u_\beta - u^\alpha a_\beta) X^\beta\ .
\eeq
This transport thus ignores the changing direction of $U$ (it has zero Fermi-Walker derivative) and only measures the rotation of the frame in the local rest space of $U$ orthogonal to the direction of the motion in spacetime.

One final remark is in order before moving on to the null case. The first Frenet-Serret equation for an arclength parametrized curve with everywhere nonvanishing curvature in a Riemannian 3-manifold with Frenet-Serret frame $\{E_1,E_2,E_3\}$, namely
\beq
  \rmD E_1/\rmd s =\kappa E_2 \rightarrow \rmD E_1/\rmd\Lambda = E_2 {\ \rm if\ } \rmd\Lambda/\rmd s =\kappa\ ,
\eeq
suggests an alternative parametrization of the world line by a turning angle parametrization, namely by the angle of rotation of the unit tangent in the osculating plane relative to projected parallel transport (most familiar in the plane curve scenario in $E_2$ \cite{gray}, and whose integral over a closed plane curve is the analog of the Gauss-Bonnet theorem in that setting). In terms of a general initial parametrization, $\rmd s/\rmd\lambda = v$, 
$\rmd x/\rmd\lambda = v E_1$, and $\rmd E_1/\rmd\lambda = v\kappa E_2$ so that $\rmd\Lambda/\rmd\lambda = v\kappa$ is the magnitude of the transverse acceleration if $\lambda$ is interpreted as the classical mechanical time, equal to the angular velocity of the unit tangent and unit normal in the osculating plane. From either starting point, in the new parametrization
this makes the coefficient of the unit normal $E_2$ in this Serret-Frenet equation identically 1 along the curve, apparently first introduced by Vessiot \cite{vessiot} (according to Bonnor \cite{bonnor1969}) who referred to it as the pseudo-arclength, loosely translated.
For an everywhere accelerated timelike world line in a spacetime, the corresponding situation instead leads to a boost angle (rapidity) parametrization of the world line.

In the null case, however, without a way to fix the parametrization defining 
$E_1 = \rmd x/\rmd\lambda$ by a normalization condition as can be done in the nonnull case, the curvature defined by 
$\rmD E_1/\rmd\lambda = \mathcal{K} E_2$ acquires two factors of $\rmd\Lambda/\rmd\lambda$ in its transformation, and furthermore the curvature and torsions no longer have a parametrization independent meaning, with the coefficient of $E_2$ transforming like a second derivative modulo projection instead of transforming by one factor of $\rmd\Lambda/\rmd\lambda$ in the nonnull case, where the curvature itself is already invariantly defined by a quotient of the coefficient of $E_2$ in that equation by $\rmd s/\rmd\lambda$.

\section{The Frenet-Serret frame for a null world line}

Consider a  Newman-Penrose frame along a given null parametrized world line with tangent vector $l$  consisting of two real and two complex conjugate null vector fields $e_1=l$, $e_2=n$, $e_3=m$, $e_4=\bar m$, satisfying the conditions
$l \cdot n=1$ and $m \cdot \bar m=-1$.
The components of the metric tensor $g_{ab}$, ($a=1,2,3,4$) in this frame are
\beq
(\eta_{ab}) = (\eta^{ab}) 
= \pmatrix{
0 & 1 & 0 & 0 \cr
1 & 0 & 0 & 0 \cr
0  & 0 & 0 & -1\cr
0  & 0 & -1& 0 \cr
}.
\eeq
The evolution equations for the frame vectors along the world line of $l$ are given by \cite{ChandraBook}
\begin{eqnarray}
\label{chandraeqs}
\frac{Dl}{\rmd\lambda}&=&(\epsilon+\epsilon^*)l-\kappa {\bar m}-\kappa^* m\ ,\nonumber \\
\frac{Dn}{\rmd\lambda}&=&-(\epsilon+\epsilon^*)n+\pi m+\pi^*{\bar m}\ ,\nonumber \\
\frac{Dm}{\rmd\lambda}&=&(\epsilon-\epsilon^*)m+\pi^* l-\kappa n\ .
\end{eqnarray}
Such a Newman-Penrose frame can be transformed into any other such frame by Lorentz transformations which, however, are best repackaged into three classes in terms of their action on frames of this type
\begin{description}
\item[\phantom{II}I.] 
$l \rightarrow l\,,\ 
m \rightarrow m +a l\,,\
n \rightarrow n +a^* m + a \bar m + a a^* l$ (null rotation);
\item[\phantom{I}II.]
$n \rightarrow n\,,\ 
m \rightarrow m +b l\,,\
l \rightarrow l +b^* m + b \bar m + b b^* n$ (null rotation);
\item[III.]
$l \rightarrow A^{-1}l\,,\ 
m \rightarrow e^{i\theta}m\,,\
n \rightarrow A n$ (boost and orthogonal rotation).
\end{description}

Since the arclength along a null curve is zero, there is no preferred parametrization of the given world line associated with the first derivative (tangent) of the curve, i.e., no immediate way to fix the scale of $l$, so one may use both class I null rotations fixing $l$ as well as class III boosts and rotations of the frame under which $l$ is rescaled in order that the first frame vector remain tangent to a given null world line. The obvious solution is to move on to use the length of the second derivative, assumed to be nonzero (nongeodesic motion) to get a new invariant parametrization.

One can arrange that the spin coefficient $\epsilon$ be zero by performing a class III  Lorentz transformation of the frame, under which
\beq\fl\qquad
\kappa\to \frac{e^{i\theta}}{A^2}\kappa\ , \qquad \pi\to e^{-i\theta}\pi\ , \qquad 
\epsilon\to \frac{\epsilon}A - \frac{1}{2A^2}\frac{DA}{\rmd\lambda}+\frac{i}{2A}\frac{D\theta}{\rmd\lambda}\ ,
\eeq
where $A$ and $\theta$ are two real functions. 
This fixes the parametrization up to affine transformations for a given choice of $m$, but now there is no natural way to fix the multiplicative coefficient in this affine freedom as in the nonnull case where the arclength picks out a preferred parametrization modulo additive constants. However, one can still use class I null rotations to change $m$ and $n$ which makes $\epsilon$ nonzero again, and one can then do another class III transformation to eliminate the new value, leading to a second rescaling of $l$ corresponding to a new parametrization, which is to say there simply is no natural parametrization of the null world line as in the nonnull case.

However, we can choose to calculate with a frame in which $\epsilon=0$ leading to the simpler equations
\beq
\label{chandraeqs2}
\frac{Dl}{\rmd\lambda} = -\kappa {\bar m}-\kappa^* m\ ,\
\frac{Dn}{\rmd\lambda} = \pi m+\pi^*{\bar m}\ ,\
\frac{Dm}{\rmd\lambda} = \pi^* l-\kappa n\ .
\eeq
When in addition $l$ is a geodesic ($\kappa=0$), this makes $\lambda$ an affine parameter, the closest analog of the proper time for timelike geodesics. Of course if $l$ is a geodesic, the construction of the Frenet-Serret frame breaks down, but then one can simply choose an adapted quasi-orthogonal frame along the world line, characterized only by the torsions. When a single geodesic is part of a family of nongeodesic world lines, it inherits a limiting Frenet-Serret frame by continuity, when such a limit exists.

The covariant derivative of a null vector $l$ cannot be timelike since it must be orthogonal to $l$, and hence either it is zero or proportional to $l$ or spacelike. In the first two cases $l$ is geodesic, which we then handle as described above. Thus if it is spacelike, it can be normalized, 
leading to the second Frenet-Serret frame vector and the curvature (acceleration corresponding to the fixed parametrization). Computing the covariant derivative of the second frame vector using the Newman-Penrose equations and removing a piece along $l$ defines the third frame vector, a null vector, and finally its derivative, removing the piece along the second Frenet-Serret frame vector, leads to the fourth one, defining the two torsions along the way. 

Given this initial choice of parametrization, we are led to the Frenet-Serret frame $\{E_a\}$ 
consisting of two real null vectors $E_1 $ and $E_3$ and two real orthogonal spacelike vectors $E_2$ and $E_4$.
One finds using the Newman-Penrose relations
\begin{eqnarray}
\label{FSnullframegen}
E_1&=&
  l
\ , \nonumber\\
E_2&=&
  \frac{1}{\sqrt{2\kappa\kappa^*}}(\kappa^* m+\kappa {\bar m})
  =\frac{1}{\sqrt{2}}(e^{-i\theta_{\kappa}} m+e^{i\theta_{\kappa}}{\bar m})
\ , \nonumber\\
E_3&=&n+X_{\kappa}\left[X_{\kappa}l+i(e^{-i\theta_{\kappa}} m-e^{i\theta_{\kappa}}{\bar m})\right]\ , \nonumber\\
E_4&=&\sqrt{2}X_{\kappa}l+\frac{i}{\sqrt{2}}(e^{-i\theta_{\kappa}} m-e^{i\theta_{\kappa}}{\bar m})\ ,
\end{eqnarray}
where a dot denotes ordinary differentiation with respect to the parameter $\lambda$ and the following notation is convenient
\beq\fl\qquad
\label{polarnot}
\kappa=\kappa_0{}e^{i\theta_{\kappa}}\ , \quad 
\pi=\pi_0{}e^{i\theta_{\pi}}\ , \quad 
X_{\kappa}=\dot\theta_{\kappa}/(2\kappa_0)\ , \quad 
w=\theta_{\kappa}+\theta_{\pi}\ .
\eeq
Apart from the permutation, the standard quasi-orthogonal real null frame inner product conditions have been used
\beq\fl
\label{metric_const}
(g_{ab}) =  (E_a\cdot E_b) = (-1)^{a-1} \delta_{a (4-b)_{\rm mod 4}}
= \pmatrix{
0 & 0 & 1 & 0 \cr
0 & -1 & 0 & 0 \cr
1  & 0 & 0 & 0\cr
0  & 0 & 0& -1 \cr
}=(g^{ab})\ .
\eeq
For interpretation, one can always introduce the associated orthonormal frame
\beq\fl\qquad
  e_0 = 2^{-1/2}(E_1+E_3)\ ,\ 
  e_1 = 2^{-1/2}(E_1-E_3)\ ,\
  e_2 = E_2\ ,\
  e_3 = E_4\ ,
\eeq
or its associated complex null frame
\beq
  L = E_1\ ,\ N = E_3\ ,\ M = 2^{-1/2}(E_2+ i E_4)\ .
\eeq
Note that in the simplest case $X_\kappa=0$, this transformation from the original real null frame associated with the original complex null frame only picks out a preferred direction in the plane of $m,\bar m$ by rotating the associated real orthonormal vectors in that plane, while an additional null rotation of the frame results in the general case to first tilt that plane correctly and then orient the second real null vector correctly to align them both with the differential properties of the curve.

This frame  satisfies the following set of Frenet-Serret evolution equations, following from the Newman-Penrose equations (\ref{chandraeqs2})
\begin{eqnarray}
\label{FSnulleqs}
\frac{\rmD E_1}{\rmd\lambda} &= -\mathcal{K} E_2\ , \qquad &  
\frac{\rmD E_2}{\rmd\lambda} = \mathcal{T}_1 E_1 -\mathcal{K} E_3\ ,\nonumber \\
 \nonumber \\
\frac{\rmD E_3}{\rmd\lambda} &= \mathcal{T}_1 E_2 + \mathcal{T}_2 E_4\ , \qquad &
\frac{\rmD E_4}{\rmd\lambda} = \mathcal{T}_2 E_1\ ,
\end{eqnarray} 
where the quantities 
\beq\fl
\label{FSlikeKetau12gen}
\mathcal{K} = \sqrt{2\kappa\kappa^*}=\sqrt{2}\kappa_0\ , \quad
\mathcal{T}_1 = \sqrt{2}[\kappa_0X_{\kappa}^2+\pi_0\cos{w}]\ , \quad
\mathcal{T}_2 = \sqrt{2}[\dot X_{\kappa}+\pi_0\sin{w}]
\eeq
play roles analogous to the Frenet-Serret curvature and torsions in the timelike case. The connection matrix
\beq
  \rmD E_a/ \rmd\lambda = E_b C^b{}_a
\eeq
has the components 
\beq\fl
   (C^a{}_b)
=\pmatrix{
0&\mathcal{T}_1 &0&\mathcal{T}_2\cr
-\mathcal{K}&0&\mathcal{T}_1&0\cr
0&-\mathcal{K}&0&0\cr
0&0&\mathcal{T}_2&0\cr}
\ ,\quad
   (C^{ab})
=
\pmatrix{
0&-\mathcal{T}_1 &0&-\mathcal{T}_2\cr
\mathcal{T}_1&0&-\mathcal{K}&0\cr
0&\mathcal{K}&0&0\cr
\mathcal{T}_2&0&0&0\cr}
\ .
\eeq 

The invariants of the matrix are given by
\begin{eqnarray}
I_1=\frac{1}{2}C_{ab}C^{ab}= 2\mathcal{K}\mathcal{T}_1\ , \qquad
I_2=\frac{1}{2}C_{ab}{}^*C^{ab}=2\mathcal{K}\mathcal{T}_2\ ,
\end{eqnarray}
where
\beq
({}^*C^{ab}) = (\frac12 \eta^{abcd}C_{cd})
= \pmatrix{
0&-\mathcal{T}_2&0&\mathcal{T}_1\cr 
\mathcal{T}_2&0&0&0\cr 
0&0&0&\mathcal{K}\cr
-\mathcal{T}_1 &0&-\mathcal{K}&0\cr  
}
\ . 
\eeq
One can also form the self-dual combination
\beq
{\mathcal C}_{ab}=C_{ab}+i{}^*C_{ab},
\eeq
so that the two invariants $I_1$ and $I_2$ are then combined into a complex invariant
\beq
\label{complinvar}
I=I_1+iI_2=\frac12 {\mathcal C}_{ab}{\mathcal C}^*{}^{ab}=2\mathcal{K} (\mathcal{T}_1+i\mathcal{T}_2)\ ,
\eeq
an expression which can will be useful in the case of stationary axisymmetric spacetimes below
(see Eq.~(\ref{useful})).
Finally one can evaluate the four complex eigenvalues of the matrix ($C^a{}_b$). 
Letting $\Omega_{\mathcal{T}}= (\mathcal{T}_1^2+\mathcal{T}_2^2)^{1/2}$, one finds
\beq\fl\qquad
\lambda_{1,2} =\pm \chi 
\equiv [\mathcal{K}(-\mathcal{T}_1+\Omega_{\mathcal{T}})]^{1/2}\ ,\quad
\lambda_{3,4} = \pm i \omega
\equiv i [\mathcal{K}(\mathcal{T}_1+\Omega_{\mathcal{T}})]^{1/2}\ ,
\eeq
which in turn defines two nonnegative quantities $\chi$ and $\omega$ \cite{bonnor1969}.

To analyze the separate effects of the curvature and the two torsions, consider the obvious decomposition
\beq
  C = C_{\mathcal{K}} + C_{\mathcal{T}}
\eeq
with
\beq
   C_{\mathcal{K}} = -\mathcal{K} (E_2 \otimes W^1 + E_3 \otimes W^2)\ ,\quad 
   C_{\mathcal{K}}^\sharp = -\mathcal{K}\, E_2 \wedge E_3 \ , 
\eeq
and
\beq
C_{\mathcal{T}} = [\mathcal{T}_1 (E_1 \otimes W^2 + E_2 \otimes W^3 ) 
                + \mathcal{T}_2 (E_1 \otimes W^4 +  E_4 \otimes W^3)]\ .
\eeq
Introducing a new pair of spatial unit vectors obtained by the rotation
\beq\label{E2F}
         \pmatrix{F_2 & F_4} = \pmatrix{E_2 & E_4} \,  
          \pmatrix{ \cos\Theta & -\sin\Theta\cr 
                    \sin\Theta &  \cos\Theta\cr}\ ,
\eeq
where $(\cos\Theta,\sin\Theta) = ( \mathcal{T}_1/\Omega_{\mathcal{T}},\mathcal{T}_2/\Omega_{\mathcal{T}} )$, 
then
\beq
C_{\mathcal{T}}^\sharp 
=- \Omega_{\mathcal{T}}\, E_1 \wedge F_2 \ .
\eeq
Thus both contributions $  C_{\mathcal{K}}$ and $  C_{\mathcal{T}} $ to the Lorentz generator, in completely contravariant form, are proportional to the wedge product of a null vector and a unit spacelike vector, which generate null rotations, or in equivalent language, parabolic Lorentz transformations, with a null rotation ``angular velocity" given by the coefficient.
The curvature part of the parallel transport matrix generates a family of null rotations of class II with an angular velocity $|\mathcal{K}|$ in the timelike hyperplane with unit normal $E_4$, which generalizes the osculating plane of the nonnull case to an osculating hyperplane in this null case. 
When the second torsion vanishes, this unit normal vector is parallel transported.
The torsion part of the parallel transport matrix $C_{\mathcal{T}}$ instead generates a family of null rotations with an angular velocity $\Omega_{\mathcal{T}}$ in the timelike hyperplane orthogonal to the spacelike unit normal vector $F_4$, rotated by the angle $\Theta$ relative to $E_4$. 

Note that bivectors which are the wedge product of a null vector and an orthogonal spacelike vector are called null flags, with the null vector acting as the flag pole and the spacelike vector, which is only defined modulo additive multiples of the null vector in the wedge product, acting as the flag orienting the 2-plane of the two vectors \cite{penrin}. It can be thought of as specifying a polarization for a photon along its world line.

Along a Killing trajectory the Newman-Penrose connection scalars and Frenet-Serret scalars are all constant and hence $X_{\kappa} = 0$, leading to considerable simplification of the above formulas. This is the situation for null circular orbits in stationary axisymmetric spacetimes, for example.  We discuss in detail this special but rather interesting case below by applying our results to null circular orbits in the Kerr black hole spacetime.

There remains the problem that the whole Frenet-Serret frame machinery is still subject to reparametrization.
The curvature is just the norm of the second derivative
\beq
  \mathcal{K} =\left(\frac{\rmD^2 x}{\rmd\lambda^2} \cdot \frac{\rmD^2 x}{\rmd\lambda^2}\right)^{1/2} \ge 0
\eeq
and it transforms by the square of the related rate of change of the parameters $\lambda \to \hat\lambda$, namely
\beq
   \mathcal{K} = \hat \mathcal{K} \, \left( \frac{\rmd\hat\lambda}{\rmd\lambda} \right)^2\ .
\eeq
Thus $\rmd\Lambda = \mathcal{K}^{1/2} \rmd\lambda = \hat \mathcal{K}{}^{1/2} \rmd\hat\lambda$ is invariant (when nonzero), like the arclength in the nonnull case, and can be used to introduce a preferred unit curvature parametrization when the curvature is everywhere nonvanishing, defined up to an additive constant like the arclength in the null case. This was already introduced by Vessiot for the Riemannian case in the same miraculous year 1905 that special relativity was proposed \cite{vessiot}, who called $\Lambda$ the pseudo-arclength.

The full transformation of all the Frenet-Serret quantities is straightforwardly calculated. Letting 
$\Lambda^\prime = \rmd \Lambda/\rmd\lambda$ and using a hat notation for the new quantities, one finds the frame vectors undergo a boost and a null rotation, with the invariance of $E_4$ confirming the invariant nature of the osculating hyperplane for which it is the unit normal
\begin{eqnarray}
\hat E_1 &= (\Lambda^\prime)^{-1} E_1\ ,\quad &
\hat E_2 =  E_2 +\zeta E_1\ ,
\nonumber\\
\hat E_3 &= \Lambda^\prime \left(E_3 + \zeta E_2 + \frac{\zeta^2}{2}E_1\right)\ ,\quad &
\hat E_4 = E_4\ ,
\end{eqnarray}
where $\zeta=\Lambda^{\prime\prime}/(\mathcal{K}\Lambda^\prime)$, and the new Frenet-Serret quantities are given by 
\beq
\hat\mathcal{K} = (\Lambda^\prime)^{-2} \mathcal{K}\ ,\quad
\hat\mathcal{T}_1 = \mathcal{T}_1+\zeta^\prime +\frac{\mathcal{K}}{2}\zeta^2\ ,\quad
\hat\mathcal{T}_2 = \mathcal{T}_2 \ .
\eeq
The pseudo-arclength parametrization is obtained by setting $\Lambda^\prime = \mathcal{K}^{1/2}$ (so that $\zeta=\mathcal{K}^\prime/(2\mathcal{K}^2)$), and the new form of the Frenet-Serret equations, once hats are dropped,  can be obtained from (\ref{FSnulleqs}) simply by making the substitutions  $\mathcal{K}\to1$ and $\lambda\to\Lambda$.
Note that in the case of constant curvature $\mathcal{K}$, then $\Lambda^{\prime\prime}=0$ (and $\zeta=0$) and this is just a simple affine parameter transformation and both torsions remain unchanged, while the frame undergoes a simple boost constant rescaling.

\section{Fermi-Walker transport along a null world line}

Exactly as in the case of a timelike curve, one may retain only the torsion part of the parallel transport matrix while absorbing the curvature part into the derivative to define a corresponding generalized Fermi-Walker transport along the null world line \cite{castagnino1965}, defined by
\beq\fl\qquad
  \rmD_{\rm(fw)} X^\alpha/\rmd\lambda 
= \rmD X^\alpha/\rmd\lambda - C_{\mathcal{K}}{}^\alpha{}_\beta X^\beta 
= \rmD X^\alpha/\rmd\lambda  +{\mathcal K}[E_2\wedge E_3]^{\alpha}{}_\beta X^\beta
= 0\ .
\eeq
This removes the piece of $C$ responsible for Lorentz transforming $l$ compared to its parallel transported direction allowing $l$ to remain tangent to the world line under the new transport:
$ \rmD_{\rm(fw)} l/\rmd\lambda =0$.
However, this extra piece cannot be expressed in terms of the tangent $l$ and its derivative as in the timelike case.
This derivative effectively acts on the quotient space of the tangent space by $l$, since $\rmD_{\rm(fw)} l/\rmd\lambda = 0$, adding multiples of $l$ to a vector does not change its Fermi-Walker derivative. 
Only the two torsions are left to describe the Fermi-Walker transport
\beq\fl\qquad
\frac{\rmD_{\rm (fw)}E_2}{\rmd \lambda } = {\mathcal T}_1 E_1\ ,\
\frac{\rmD_{\rm (fw)}E_3}{\rmd \lambda } = {\mathcal T}_1 E_2+{\mathcal T}_2 E_4\ ,\
\frac{\rmD_{\rm (fw)}E_4}{\rmd \lambda } = {\mathcal T}_2 E_1\ .
\eeq

In the nonnull case, one can use orthogonality to pick the orthogonal complement as a preferred representative subspace for the quotient space by the unit tangent vector, but in this null case, orthogonality does not lead to a complementary subspace but contains $l$ itself. 
Samuel and Nityananda \cite{samraj2000}
instead confine their attention to the orthogonal complement and in turn to its quotient by $l$, ignoring the fourth direction, and are led to the restriction of this Fermi-Walker transport to that subspace where it is possible to express the Fermi-Walker add-on in terms of the first two derivatives
\begin{eqnarray}\fl
 C_{\rm(SN)}{}^\sharp &=& 
                         \frac{\rmD^2 l}{\rmd\lambda^2} \wedge \frac{\rmD l}{\rmd\lambda} 
                       / \left(\frac{\rmD l}{\rmd\lambda} \cdot \frac{\rmD l}{\rmd\lambda} \right)
                       =[\mathcal{T}_1  E_1-\mathcal{K} E_3 ]\wedge E_2 =-[C_\mathcal{K}{}^\sharp +\mathcal{T}_1 E_2\wedge E_1] \ .
\end{eqnarray}
If one ignores additive multiples of $l$, the extra term does not contribute to the derivative defined for equivalence classes of vectors mod $l$. They were only interested in defining a derivative for polarization vectors along null world lines, and these can be identified with this quotient of the orthogonal complement. These spacelike vectors modulo additive multiples of $l$ define a null flag.

In the special case in which $l$ is a null Killing vector field, the curvature and torsions are constant along the curve and a Fermi-Walker transported quasi-orthogonal frame $\{H_a = E_b P^b{}_a\}$ is easily constructed. 
The parallel transport condition is just 
\beq
0=\frac{\rmD H_a}{\rmd\lambda} = E_b ( \frac{\rmd P^b{}_a}{\rmd\lambda} +P^b{}_c C^c{}_a)
\eeq
and with the initial condition $P(0)=I$, the solution is just $P(\lambda) = \exp(-\lambda C_{\mathcal{T}} )$, namely
\beq\label{eq:EHTT}
\fl
 \pmatrix{H_1 & H_2 & H_3 & H_4\cr}
= \pmatrix{E_1 & E_2 & E_3 & E_4\cr}
  \pmatrix{ 1 & -\lambda \mathcal{T}_1 & {\textstyle\frac12}\lambda^2 \Omega_{\mathcal{T}}{}^2 &
                -\lambda \mathcal{T}_2\cr
         0 & 1 &  -\lambda \mathcal{T}_1 & 0\cr
         0 & 0 &  1 & 0\cr
         0 & 0 &  -\lambda \mathcal{T}_2 & 1\cr } \ . 
\eeq
This transformation is a one-parameter family of null rotations in a hyperplane whose normal is a fixed rotation of the normal to the osculating hyperplane in the plane of the two spacelike frame vectors $E_2$ and $E_4$
\beq\fl
\pmatrix{1 & 0 & 0 & 0\cr
         0 & \cos\Theta & 0 & -\sin\Theta \cr
         0 & 0   & 1 & 0\cr
         0 & \sin\Theta & 0 & \cos\Theta \cr}
\pmatrix{ 1 & -\lambda \Omega_{\mathcal{T}} & {\textstyle\frac12}\lambda^2 \Omega_{\mathcal{T}} & 0\cr
         0 & 1 &  -\lambda \Omega_{\mathcal{T}} & 0\cr
         0 & 0 &  1 & 0\cr
         0 & 0 &  0 & 1\cr }
\pmatrix{1 & 0 & 0 & 0\cr
         0 & \cos\Theta & 0 & \sin\Theta \cr
         0 & 0   & 1 & 0\cr
         0 & -\sin\Theta & 0 & \cos\Theta \cr}\ .
\eeq
where as above $(\cos\Theta,\sin\Theta) = ( \mathcal{T}_1/\Omega_{\mathcal{T}},\mathcal{T}_2/\Omega_{\mathcal{T}} )$.

\section{The equivalence classes of uniformly accelerated null world lines in Minkowski spacetime}

In the nonnull case, Killing trajectories in either $E_2$ or $M_2$ give us the simplest constant curvature curves to compare with the curvature at a point of any nonnull curve in a higher dimensional manifold, defining the osculating plane in the tangent space spanned by the tangent and second derivative and containing the osculating circle or pseudocircle (hyperbola) respectively.
By examining null Killing trajectories in Minkowski spacetime, we gain some intuition about what the Frenet-Serret quantities tell us about the geometry of such world lines for comparison with general world lines in any spacetime.

These curves have a constant Lorentz generating matrix $(C^a{}_b)$, and so may be classified by its invariants
$I_1+iI_2=2\mathcal{K} (\mathcal{T}_1+i\mathcal{T}_2)$. Null geodesics have $\mathcal{K}=0$ and are straight lines in Minkowski spacetime.
For nongeodesics, the general case is $\mathcal{T}_2\ne0$, the semisingular case is $\mathcal{T}_1\ne0,\mathcal{T}_2=0$, and the singular case is $\mathcal{T}_1=0=\mathcal{T}_2$. Bonnor has called the latter case curves  null cubics and the remaining curves null helices \cite{bonnor1969}, giving representative curves of each of the four types in the unit curvature parametrization.
All of these curves are particular integral curves belonging to a Killing vector field conguence.

Note that when $\mathcal{T}_2=0$, the spacelike vector $E_4$ is covariant constant along the curve. In Minkowski space this means that the curve is confined to a timelike hyperplane orthogonal to this vector.

Apart from isometries, any such curves are reducible to one of the following examples.

\def\Item#1:{\vskip 12pt\noindent{\bf #1}:}
\Item Generalized null helical orbits:

The parametrized equations of the generalized null helical orbits through the origin
at $\lambda=0$ in
the unit curvature parametrization (assuming $\omega^2+\chi^2\neq0$) are given by
\begin{eqnarray}
(x,y) &=& (\omega^2+\chi^2)^{-1/2}(\omega^{-1}(\cos\omega\lambda-1),\omega^{-1}\sin\omega\lambda)\ ,   
\nonumber\\ 
(z,t) &=& (\omega^2+\chi^2)^{-1/2} (\chi^{-1}(\cosh\chi\lambda-1),\chi^{-1}\sinh\chi\lambda)\ .
\end{eqnarray}
with corresponding null tangent
\beq\fl\quad
\label{helices}
l=(\omega^2+\chi^2)^{-1/2}[\cosh\chi\lambda\,\partial_t-\sin\omega\lambda\,\partial_x
                    +\cos\omega\lambda\,\partial_y+\sinh\chi\lambda\,\partial_z]\ ,
\eeq
whose initial value is 
\beq
l(0) = (\omega^2+\chi^2)^{-1/2}[\partial_t + \partial_y]\ .
\eeq
The associated Frenet-Serret curvature and torsions are
\beq
\mathcal{K}=1\ , \quad \mathcal{T}_1=\frac12(\omega^2-\chi^2)\ , \quad \mathcal{T}_2=-\chi\omega\ .
\eeq

Letting $(x_0,z_0)=(\omega^2+\chi^2)^{-1/2} (\omega^{-1},\chi^{-1})$ be the radii of the circle and pseudocircle associated with the rotation and boost respectively,
this curve is an integral curve of the Killing vector field
\beq
  \xi = \chi[(z-z_0) \partial_t + t \partial_z] 
      + \omega[(x-x_0) \partial_y - y \partial_x] 
\eeq
in the hypersurface $\chi^2[(z-z_0)^2 -t^2] -\omega^2[ (x-x_0)^2+y^2] =0$ where the vector field is null and is confined to the 3-space cylinder
$ (x-x_0)^2+y^2 = x_0^2$, around which it wraps with uniform acceleration in the $z$-direction if $\chi\neq0$. 

The Poincar\'e group generating vector field $\xi$ consists of a translation generator and a Lorentz transformation generator of the form $M(\xi)^\alpha{}_\beta x^\beta \partial_\alpha$ in terms of the inertial coordinates $(x^\alpha)=(t,x,y,z)$. The matrix $M(\xi)$ generates a
1-parameter family of Lorentz transformations 
$ e^{\lambda\,M(\xi)}$ and is explicitly
\beq\fl
   (M(\xi)^\alpha{}_\beta) =\pmatrix{0&0&0&\chi\cr 0&0&-\omega&0\cr 0&\omega&0&0\cr \chi&0&0&0\cr}\ ,\quad
   M(\xi)^\sharp=\chi(\partial_t\wedge\partial_z)+\omega(\partial_x\wedge\partial_y)\ ,
\eeq
\beq
   \left(\left[e^{\lambda\,M(\xi)}\right]^\alpha{}_\beta\right) 
    =
   \pmatrix{\cosh\chi\lambda&0&0&\sinh\chi\lambda\cr 0&\cos\omega\lambda&-\sin\omega\lambda&0\cr 0&\sin\omega\lambda&\cos\omega\lambda&0\cr \sinh\chi\lambda&0&0&\cosh\chi\lambda\cr}\ ,
\eeq
representing a Lorentz boost acting in the $t$-$z$ plane together with a rotation in the $x$-$y$ plane. Applying this matrix to the inertial coordinate component vector $(\omega^2+\chi^2)^{-1/2}(1,0,1,0)$ of $l(0)$ yields $l(\lambda)$.
This case contains the special cases of a pure boost ($\omega=0$), a pure rotation ($\chi=0$) or an equally weighted combination $(\chi=\pm\omega$) where $\mathcal{T}_1=0$. The pure rotation case corresponds to a true helix in spacetime.

\Item
Null cubics:

The parametrized equations of a null cubic curve through the origin at $\lambda=0$ in
the unit curvature parametrization are given by
\beq
(t,x,y,z)
= \left(\frac{1}{\sqrt{2}}\left(\lambda+\frac{\lambda^3}6\right),
  -\frac{\lambda^2}2,
  \frac{1}{\sqrt{2}}\left(\lambda-\frac{\lambda^3}6\right),
  0\right)\ ,
\eeq
with corresponding null tangent
\beq
\label{nullrot}
l = \frac{1}{\sqrt{2}}\left(1+\frac{\lambda^2}2\right)\,\partial_t
    -\lambda\,\partial_x
    +\frac{1}{\sqrt{2}}\left(1-\frac{\lambda^2}2\right)\,\partial_y
\ ,
\eeq
whose initial value is
\beq
  l(0) =  \frac{1}{\sqrt{2}}\,[\partial_t + \partial_y]\ .
\eeq
The associated Frenet-Serret curvature and torsions are
\beq
\mathcal{K}=1\ , \quad \mathcal{T}_1=0=\mathcal{T}_2\ .
\eeq

This curve is an integral curve of the Killing vector field
\beq
\xi = 2^{-1/2}[ \partial_t +\partial_y - (x\partial_t + t \partial_x)+(-y\partial_x+x\partial_y)]
\eeq
on the hypersurface $4x +(t+y)^2=0$, where it generates a simultaneous null translation and null rotation.
This 1-parameter family of Poincar\'e transformations contains a 1-parameter family of Lorentz transformations 
$ e^{\lambda\,M(\xi)}$
generated by $\xi$ and is itself generated by the matrix with respect to the inertial coordinates $(t,x,y,z)$
(with associated tensor)
\beq\fl
   (M(\xi)^\alpha{}_\beta) =\frac{1}{\sqrt{2}}\pmatrix{0&-1&0&0\cr -1&0&-1&0\cr 0&1&0&0\cr 0&0&0&0\cr}\ ,
 \quad M(\xi)^\sharp =-\frac{1}{\sqrt{2}}(\partial_t+\partial_y)\wedge\partial_x\ ,
\eeq
\begin{eqnarray}\fl\quad
   \left(\left[e^{\lambda\,M(\xi)}\right]^\alpha{}_\beta\right) 
  &=& \left(\left[{\rm I}+\lambda M(\xi) +{\textstyle\frac12} \lambda^2 M(\xi)^2\right]^\alpha{}_\beta\right)
\nonumber\\   \fl\quad
   &=&\pmatrix{1+\lambda^2/4&-\lambda/\sqrt{2}&\lambda^2/4&0\cr 
               -\lambda/\sqrt{2}&1&-\lambda/\sqrt{2}&0\cr 
               -\lambda^2/4&\lambda/\sqrt{2}&1-\lambda^2/4&0\cr 
               0&0&0&1\cr}\ .
\end{eqnarray}
Applying this matrix 
to the component vector of $l(0)$ in the inertial coordinates produces $l(\lambda)$.
The null cubics are a Lorentzian realization of the twisted cubic curves familiar from the multivariable calculus parametrized curve example $(x,y,z)=(t,t^2,t^3)$, except that now the quadratic behavior occurs in a null 2-plane (linear behavior along $l=E_1$ and quadratic behavior along $D l/d\lambda$ or equivalently $E_2$) with the cubic behavior along $D^2 l/d\lambda^2$ departing from this 2-plane within the osculating hyperplane \cite{urbantke1989}. These curves only exhibit curvature with vanishing torsions, and since the curvature value is not geometric, there is no additional information left for interpretation like a radius of curvature as in the nonnull case. 

However, one can compare this curve with the previous case of the generalized null helix for small values of the parameter $\lambda$ and with vanishing first torsion $\chi=\omega$. The curve itself is
\beq\fl\qquad
(t,x,y,z)
= \frac{1}{\sqrt{2}\chi^2} 
 \left(\chi\lambda+\frac{(\chi\lambda)^3}6,
      -\frac{(\chi\lambda)^2}2,
  \chi\lambda-\frac{(\chi\lambda)^3}6,
  \frac{(\chi\lambda)^2}2\right)\ .
\eeq
Apart from a 45 degree rotation $(x,z)\rightarrow 2^{-1/2}(x-z,x+z)$, this is exactly the null cubic curve above when $\chi=1$. The nonzero second torsion only affects terms higher than the third order in $\lambda$, so up to third order these curves coincide. However, by introducing new coordinates conformally rescaled by a factor $\chi^2$, one can balance the effect of rescaling the parameter $\lambda$ by $\chi$ to retain unit curvature and obtain the null cubic exactly as the limit $\chi\to0$ of the generalized null helix itself. In any case one sees that the torsion and curvature properties are mixed up in the osculating hyperplane, unlike the nonnull case. The null cubic is the simplest curve one can use to locally model the effects of curvature alone in nonflat spacetimes, as an approximating null cubic in the osculating hyperplane.

\section{Null circular orbits in stationary axisymmetric spacetimes}

Consider stationary axisymmetric spacetimes using adapted coordinates, i.e., let $\partial_t$ and $\partial_\phi$ be the timelike and spacelike Killing vectors associated with the respective symmetries. The spacetime metric has the general form
\beq
\rmd s^2 = N^2\rmd t^2 -g_{\phi \phi }(\rmd \phi+N^{\phi}\rmd t)^2 
           - g_{rr}\rmd r^2 -g_{\theta \theta}\rmd \theta^2\ ,
\eeq
where $N=(g^{tt})^{-1/2}$ and $N^{\phi}=g_{t\phi}/g_{\phi\phi}$ are the lapse function and shift vector fields.
Introduce the zero-angular-momentum observer (ZAMO) family with 4-velocity
\beq
\label{n}
e_{\hat t}=N^{-1}(\partial_t-N^{\phi}\partial_\phi)
\eeq
which is completed to an orthonormal frame by
\beq
\fl\quad
e_{\hat r}=\frac1{\sqrt{-g_{rr}}}\partial_r\ , \,\quad
e_{\hat \theta}=\frac1{\sqrt{-g_{\theta \theta }}}\partial_\theta\ , \,\quad
e_{\hat \phi}=\frac1{\sqrt{-g_{\phi \phi }}}\partial_\phi\ ,
\eeq
with dual frame
\beq
\fl\quad
\omega^{{\hat t}} = N \rmd t\ , \quad 
\omega^{{\hat r}} =\sqrt{-g_{rr}} \,\rmd r\ , \quad 
\omega^{{\hat \theta}} = \sqrt{-g_{\theta \theta }} \,\rmd \theta\ , \quad
\omega^{{\hat \phi}} = \sqrt{-g_{\phi \phi }}(\rmd \phi+N^{\phi}\rmd t)\ .
\eeq

Let $l$ be the tangent vector to a null world line in a Killing orbit cylinder corresponding to a circular orbit with unit speed
\beq
l_\pm= \Gamma_l (\partial_t + \zeta_{ ({\rm nul}, \pm)}\partial_\phi)
 =G_l[e_{\hat t} \pm  e_{\hat \phi}]\ ,
\eeq
where $\Gamma_l $ and $G_l$ are arbitrary normalization factors and the angular velocity $\zeta_{ ({\rm nul}, \pm)}$ is given by
\beq
\zeta_{ ({\rm nul}, \pm)} =-N^{\phi}\pm N / \sqrt{-g_{\phi\phi}}\ .
\eeq
The $\pm$ sign corresponds to the two possible azimuthal directions for such orbits.
One easily gets a symmetry-adapted Newman-Penrose frame by defining
\beq\fl\qquad
  n_\pm = \Gamma_n (\partial_t + \zeta_{ ({\rm nul}, \mp)}\partial_\phi)
 =G_n[e_{\hat t} \mp  e_{\hat \phi}]\ ,\quad
  m = (e_{\hat r} + i e_{\hat\theta} )/\sqrt{2} \ ,
\eeq
with $\Gamma_n=(2\Gamma_l N^2)^{-1}$ and $G_n=(2G_l)^{-1}$. 

Since this Newman-Penrose frame is Lie dragged along $l_\pm$, the phase $\theta_\kappa$ of Eq.~(\ref{FSnullframegen}) is constant along its integral curves and the $\lambda$-derivative quantity $X_\kappa$ is therefore zero, simplifying the general formulas for the Frenet-Serret frame Eq.~(\ref{FSnullframegen}) by eliminating the $X_\kappa$ terms, and those of Eq.~(\ref{FSlikeKetau12gen}) for the curvature and torsions then similarly simplify to
\beq
\label{FSlikeKetau12}
{\mathcal K}
=\sqrt{2}\kappa_0
\ ,\
{\mathcal T}_1
=\sqrt{2}\pi_0\cos{w}
\ ,\
{\mathcal T}_2
=\sqrt{2}\pi_0\sin{w}\ .
\eeq
Note that in this particular case the complex invariant (\ref{complinvar}) turns out to be simply $I=4\kappa \pi$. 

\section{Null circular orbits in Kerr spacetime}

The Kerr metric  in standard Boyer-Lindquist coordinates is given by
\begin{eqnarray}
\rmd s^2 &=& \left(1-\frac{2Mr}{\Sigma}\right)\rmd t^2 +\frac{4aMr}{\Sigma}\sin^2\theta\rmd t\rmd\phi- \frac{\Sigma}{\Delta}\rmd r^2 -\Sigma\rmd \theta^2\nonumber\\
&&-\frac{(r^2+a^2)^2-\Delta a^2\sin^2\theta}{\Sigma}\sin^2 \theta \rmd \phi^2\ ,
\end{eqnarray}
where $\Delta=r^2-2Mr+a^2$ and $\Sigma=r^2+a^2\cos^2\theta$; here $a$ and $M$ are the specific angular momentum and total mass. The event horizons are located at $r_\pm=M\pm\sqrt{M^2-a^2}$.

The symmetry-adapted Newman-Penrose frame along generic co-rotating and counter-rotating circular null orbits of the previous section, setting the normalization factor $\Gamma_l=1$, is then explicitly
\begin{eqnarray}
\fl\quad
l_\pm &=& \partial_t + \frac{2aMr\sin \theta \pm\sqrt{\Delta}\Sigma}{\sin \theta [(r^2+a^2)\Sigma +2a^2Mr\sin^2 \theta]} \partial_\phi \nonumber \\
\fl\quad
n_\pm &=& \frac{(r^2+a^2)\Sigma +2a^2Mr\sin^2 \theta}{2\Sigma \Delta}\left[ \partial_t + \frac{2aMr\sin \theta \mp\sqrt{\Delta}\Sigma}{\sin \theta [(r^2+a^2)\Sigma +2a^2Mr\sin^2 \theta]} \partial_\phi\right]\nonumber \\
\fl\quad
m &=& \frac{1}{\sqrt{2\Sigma}}\left( \sqrt{\Delta}\partial_r +i \partial_\theta\right).
\end{eqnarray}
The $l_\pm$ congruence  contains the null circular geodesic orbits in the equatorial plane not far from the outer horizon.

One finds that
\beq
\label{pisol}
\pi=\frac{1}{2\sqrt{2}\sqrt{\Sigma}}\left( \frac{r-M}{\sqrt{\Delta}}-i\, \cot\theta \right)
\eeq
for both signs $\pm$ and
\begin{eqnarray}\fl\quad
\label{kappasol}
{\rm Re}[\kappa_\pm]
&=& -\left(\frac{\Delta}{2\Sigma}\right)^{1/2}[\Delta\Sigma+2Mr(r^2+a^2)]^{-2}\nonumber\\
\fl\quad
&&\bigg\{
\Delta[-\Sigma(\Sigma(r-M)+2Mr^2)+4Mr^2(r^2+a^2)]-2M\Sigma[-Mr(3r^2+a^2)\nonumber\\
\fl\quad
&&+(r^2+a^2)^2]
\mp2Ma\sqrt{\Delta}[2r^2(r^2+a^2)+\Sigma(r^2-a^2)]\sin\theta
\bigg\}\ ,\nonumber\\
\fl\quad
{\rm Im}[\kappa_\pm]&=&\frac{\Delta}{\sqrt{2\Sigma}}[\Delta\Sigma+2Mr(r^2+a^2)]^{-2}\cot\theta\nonumber\\
\fl\quad
&&\bigg\{
\Delta\Sigma^2+2Mr(r^2+a^2)[2(r^2+a^2)-\Sigma]
\mp4Mra^3\sqrt{\Delta}\sin^3\theta
\bigg\}\ .
\end{eqnarray}

Explicit expressions for the Frenet-Serret curvature and torsions as functions of the orbital parameters $r/M$ and $\theta$ follow from Eqs.~(\ref{pisol}) and (\ref{kappasol}) together with Eq.~(\ref{FSlikeKetau12gen}) but are not very enlightenling. 
Instead we specialize to
the equatorial plane $\theta=\pi/2$ where there exist two circular null geodesics at radii 
\beq
r_{\rm (geo) \pm }
=2M\left\{ 1+\cos\left[ \frac23 {\rm arccos} \left( \pm \frac{a}{M}\right) \right] \right\}\ ,
\eeq
which are real solutions of the cubic equation
\beq
r^3-6Mr^2+9M^2r-4Ma^2=0\ .
\eeq
In the equatorial plane one finds ${\rm Im}[\pi]=0={\rm Im}[\kappa_\pm]$, so the Frenet-Serret curvature and torsions are explicitly
\begin{eqnarray}\fl\quad
{\mathcal K}&=&\sqrt{2}|{\rm Re}[\kappa_\pm]|
=\sqrt{\Delta}\left|\frac{r^3-6Mr^2+9M^2r-4Ma^2}{(r^3-3Ma^2)\Delta -M(r^4-a^4) \mp2Ma\sqrt{\Delta}(3r^2+a^2)}\right|\ , \nonumber\\
\fl\quad 
{\mathcal T}_1 
&=& \frac{{\rm Re}[\pi]}{\sqrt{2}}\,{\rm sgn}(r-r_{\rm (geo) \pm })
=\frac{r-M}{2r\sqrt{\Delta}}\,{\rm sgn}(r-r_{\rm (geo) \pm })\ ,\quad 
{\mathcal T}_2 \equiv0\ .
\end{eqnarray}
The curvature ${\mathcal K}$ vanishes only in the case of equatorial null circular orbits occurring at $\theta=\pi/2$ and $r=r_{\rm (geo) \pm}$ as expected. In addition the second torsion ${\mathcal T}_2$ is identically zero in the equatorial plane for every value of the radius, as is the case for timelike circular orbits there \cite{circfs}. 
Nonequatorial orbits are instead characterized by nonvanishing curvature and torsions; there exists at most one orbit for $r_+<r<r_{\rm (geo) \pm}$ at fixed $\theta$ where the first torsion ${\mathcal T}_1$ is allowed to vanish. 

\subsection{Schwarzschild spacetime limit}

In the limit of the Schwarzschild spacetime all these results simplify. The Newman-Penrose frame adapted to the null circular orbits is
\begin{eqnarray}\fl\qquad
l_\pm &=& \partial_t \pm\frac{\sqrt{r(r-2M)}}{r^2\sin \theta} \partial_\phi
\ ,\quad
n_\pm = \frac{r}{2(r-2M)}\left[ \partial_t \mp \frac{\sqrt{r(r-2M)}}{r^2\sin \theta} \partial_\phi\right]
\ ,\nonumber \\\fl\qquad
m &=& \frac{1}{\sqrt{2}}\left( \sqrt{1-\frac{2M}{r}}\partial_r +\frac{i}{r} \partial_\theta\right)\ ,
\end{eqnarray}
the spin coefficients $\pi$ and $\kappa$ for both signs are
\begin{eqnarray}
\pi&=&\frac{1}{2\sqrt{2}r}\left[ \frac{r-M}{\sqrt{r(r-2M)}}-i\, \cot\theta \right]\ , \nonumber\\
\kappa&=&\frac{\sqrt{2}}2\frac{\sqrt{r(r-2M)}}{r^3}[r-3M+i\, \sqrt{r(r-2M)}\cot\theta]\ ,
\end{eqnarray}
while the Frenet-Serret curvature and torsions (\ref{FSlikeKetau12}) are
\begin{eqnarray}
\label{FSlikectsschw}
{\mathcal K}&=&\frac{\sqrt{r(r-2M)}}{r^3\sin\theta}[(r-3M)^2+M(4r-9M)\cos^2\theta]^{1/2}
\ , \nonumber \\
{\mathcal T}_1&=&\frac1{2{\mathcal K}}\frac1{r^4\sin^2\theta}[(r-M)(r-3M)-M(3M-2r)\cos^2\theta]
\ , \nonumber \\
{\mathcal T}_2&=&\frac{M}{{\mathcal K}r^4}\sqrt{r(r-2M)}\cot\theta\ .
\end{eqnarray}
Note that the curvature vanishes only at $r=3M$, $\theta=\pi/2$ corresponding to null geodesic orbits; the first torsion instead vanishes at 
\beq
r=M\left[2-\cos^2\theta\pm\sqrt{1-\sin^2\theta\cos^2\theta}\right]
\eeq
ranging from $2M$ to $3M$; the second torsion vanishes identically only in the equatorial plane.
In particular, in the equatorial plane these expressions reduce to
\beq\fl\quad 
{\mathcal K}=\frac{\sqrt{r(r-2M)}}{r^3}|r-3M|\ , \quad 
{\mathcal T}_1=\frac{1}{2r}\frac{r-M}{\sqrt{r(r-2M)}}\,{\rm sgn}(r-3M)\ , \quad 
{\mathcal T}_2\equiv0\ .
\eeq

\subsection{Flat spacetime limit}

In the Minkowski spacetime limit (i.e., Schwarzschild with $M=0$) further simplifications occur
\beq\fl\quad
\label{NPmink}
l_\pm = \partial_t \pm \frac{1}{r\sin \theta} \partial_\phi\ , \quad
n_\pm = \frac12\left[ \partial_t \mp \frac{1}{r\sin \theta} \partial_\phi\right]\ , \quad
m = \frac{1}{\sqrt{2}}\left( \partial_r +\frac{i}{r} \partial_\theta\right)\ ;
\eeq
the spin coefficients $\pi$ and $\kappa$ for both signs are given by
\beq
\pi=\frac{1}{2\sqrt{2}r}(1-i\, \cot\theta)\ , \qquad
\kappa=\frac{1}{\sqrt{2}r}(1+i\, \cot\theta)=2\pi^*\ ,
\eeq
so the Frenet-Serret curvature and torsions (\ref{FSlikeKetau12}) are
\beq
{\mathcal K}=\frac{1}{r\sin\theta}={\mathcal T}_1\ , \qquad
{\mathcal T}_2\equiv0\ .
\eeq
Clearly there are no circular null geodesics.

Thse results are much easier understood in standard cylindrical coordinates $\{t,\rho,\phi,z\}$, in terms of which the line element is
\beq
\rmd s^2 = \rmd t^2  -\rmd \rho^2 -\rho^2\rmd \phi^2-\rmd z^2\ .
\eeq
The Newman-Penrose frame can be chosen such that
\beq\fl\quad
l_\pm = \partial_t \pm \partial_{\hat \phi}\ , \quad
n_\pm = \frac12(\partial_t \mp \partial_{\hat \phi})\ , \quad
m = \frac{1}{\sqrt{2}}\left( \partial_\rho -i\partial_z\right)\ .
\eeq
It differs from the corresponding one obtained by transforming (\ref{NPmink}) only by an overall factor in the definition of $m$ (and $\bar{m}$).

The relevant spin coefficients\footnote{Note that this slightly different choice of the Newman-Penrose frame leaves the spin coefficient $\epsilon=0$, as required.}
are $\kappa=1/(\sqrt{2}\rho)=2\pi$
so the curvature and torsions (\ref{FSlikeKetau12}) are
${\mathcal K}=\frac{1}{\rho}={\mathcal T}_1$ and
${\mathcal T}_2\equiv0$.
The Frenet-Serret frame simplifies to
\beq
E_1=l_\pm\ , \quad E_2=\partial_\rho\ , \quad E_3=n_\pm\ , \quad E_4=\partial_z\ .
\eeq 

\section{Quasi-orthogonal frames parallel propagated along a null geodesic}

Marck \cite{marck1} elegantly solved the problem of determining a natural quasi-orthogonal frame $\{e_a\}$ parallel transported along a null geodesic with tangent $l$ in the Kerr and Kerr-Newman spacetimes, which admit a Killing-Yano antisymmetric tensor $a_{\alpha\beta}$ of rank $2$ defined by the condition
$
a_{\alpha (\beta; \gamma)}=0$.

Let $l=e_1$ denote the first vector of the frame; the second one $e_2$ can be chosen to be
$e_2^\alpha =a^\alpha{}_{\beta}e_1^\beta$, which in general is orthogonal to $e_1$ because of the antisymmetry of $a^\alpha{}_{\beta}$.
It is also parallel transported along $e_1$ due to the antisymmetry of the Killing-Yano tensor
\beq
\nabla_{e_1}e_2^\alpha = a^\alpha{}_{(\beta; \gamma)}e_1^\beta e_1^\gamma=0\ .
\eeq
The remaining two vectors $e_3$ and $e_4$ are then obtained first by completing the first two vectors to a quasi-orthogonal frame easily determined from the symmetry and geodesic conditions, and then
integrating the simplified transport equations in the timelike hyperplane orthogonal to $e_2$, which determine a family of class I null rotations of these last two vectors by a null rotation angle expressible through elliptic integrals. He later extended the technique to the case of a timelike or spacelike vector in a spacetime admitting a Killing-Yano tensor satisfying some additional conditions \cite{marck2}. 
 
However, Marck's approach breaks down when $l$ is a principal direction of the Killing-Yano tensor itself, because in that case $a^\alpha{}_{\beta}e_1^\beta \propto e_{1}^\alpha$ and it does not determine a linearly independent vector.
For the Kerr or Kerr-Newman spacetimes, for example,
all the relevant spacetime fields, namely the Weyl tensor, the electromagnetic field tensor (if present), the Papapetrou field associated with the timelike Killing vector, and the Killing-Yano 2-tensor, all have the same principal null directions, and his approach cannot be used to find a parallel transported frame along these directions.

However,  in any spacetime for which one can integrate the equations for null geodesics explicitly, one can also directly integrate the equations of parallel transport along such null geodesics when expressed in Frenet-Serret form, which is apparently a new result. This process is essentially equivalent to generalizing the integration of the case of constant Frenet-Serret scalars by exponentiation of the parallel transport matrix to an actual step-by-step integration process which produces almost the same result apart from a quadrature.

Let $\lambda$ be an affine parameter for the world line of $l$ and complete it to any adapted quasi-orthogonal frame  $\{E_a\}$ along it (with zero curvature clearly).
The equations $\rmD W/\rmd\lambda=0$ of parallel transport for a vector $W=W^a E_a$ are
\begin{eqnarray}&&
\left(\frac{\rmd W^1}{\rmd\lambda} + \mathcal{T}_1 W^2 + \mathcal{T}_2 W^4\right) E_1
 +\left(\frac{\rmd W^2}{\rmd\lambda} + \mathcal{T}_1 W^3\right) E_2 \nonumber\\&&\qquad
 +\frac{\rmd W^3}{\rmd\lambda} E_3
 +\left(\frac{\rmd W^4}{\rmd\lambda} + \mathcal{T}_2 W^3\right) E_4
=0\ .
\end{eqnarray}
The third component must be a constant $W^3(\lambda) =W^3(\lambda_0)$, which then allows $W^2$ and $W^4$ to be integrated by introducing $T_1 =\int_{\lambda_0}^\lambda \mathcal{T}_1 \, \rmd\lambda$ and 
$T_2 =\int_{\lambda_0}^\lambda \mathcal{T}_2 \, \rmd\lambda$, so that
\beq\fl
  W^2(\lambda) = W^2(\lambda_0)- T_1(\lambda) W^3(\lambda_0)\ ,\
  W^4(\lambda) = W^4(\lambda_0)- T_2(\lambda) W^3(\lambda_0)\ .
\eeq
Finally the remaining component is easily obtained
\beq\fl
  W^1(\lambda) = W^1(\lambda_0) - T_1(\lambda) W^2(\lambda_0) - T_2(\lambda) W^4(\lambda_0)
                   + {\textstyle\frac12} (T_1(\lambda){}^2+ T_1(\lambda){}^2) W^3(\lambda_0)\ .
\eeq
This result takes the form $W^a(\lambda) = P^a{}_b(\lambda) W^b(\lambda_0)$ with
\beq
   (P^a{}_b) = 
\pmatrix{ 
1 & -T_1 & {\textstyle\frac12} (T_1{}^2 + T_2{}^2) & - T_2\cr
0 & 1 & -T_1 & 0\cr
0 & 0 & 1 & 0\cr
0 & 0 & -T_2 & 1\cr} \ .
\eeq
Since setting $W^a(\lambda_0) = \delta^a{}_b$ for a given value of $b$ corresponds to the initial data $W(\lambda_0)=E_b(\lambda_0)$, one finds the parallel transported components are $P^a{}_b(\lambda)$, so defining 
$H_a = E_b P^b{}_a$ along the curve shows that the columns of this matrix are the components of a parallel transported frame which is quasi-orthogonal since the initial frame is quasi-orthogonal and parallel transport preserves inner products. The parallel transported frame is
\begin{eqnarray}\fl\qquad
H_1 &=&  E_1\ ,\qquad
H_2 =  E_2 - T_1 E_1\ ,\nonumber\\\fl\qquad
H_3 &=&  E_3 + {\textstyle\frac12} (T_1{}^2 + T_2{}^2) E_1 - T_1 E_2 -T_2 E_4\ ,\qquad
H_4 =  E_4 - T_2 E_1\ .
\end{eqnarray}
When the torsions are constant, then $T_1=\mathcal{T}_1 (\lambda-\lambda_0)$ and $T_2=\mathcal{T}_2 (\lambda-\lambda_0)$ and setting $\lambda_0=0$, this is exactly the result Eq.~(\ref{eq:EHTT}) for Fermi-Walker transport in the Killing trajectory case. The interpretation is the same except that the rotation which re-orients the normal to the hyperplane of the null rotation relative to the Frenet-Serret frame now depends on $\lambda$.

In the Kerr spacetime, using standard Boyer-Lindquist coordinates, the geodesic and shearfree principal null direction is given by:
\beq
E_1 =l=\frac{r^2+a^2}{\Delta}\partial_t + \partial_r + \frac{a}{\Delta}\partial_\phi
\eeq
and an affine parameter along $l$ is $\lambda=r-r_+ +\lambda_0$, with $r_+=M+\sqrt{M^2-a^2}$ the outer horizon radius;
conveniently we set  $\lambda_0=r_+$ so that $\lambda=r$.
The standard Newman-Penrose frame introduced by Kinnersley \cite{kinnersley} and specialized by Teukolsky \cite{teukolsky} and the associated real quasi-orthogonal frame are
\begin{eqnarray}
\qquad E_3 &=& n = \frac{\Delta}{2\Sigma}
\left( \frac{r^2+a^2}{\Delta}\partial_t - \partial_r + \frac{a}{\Delta}\partial_\phi \right)\ ,
\nonumber \\
E_2+ i E_4 &=& \sqrt{2} m = \frac{1}{(r+ia \cos \theta)}\left( ia\sin \theta \partial_t +\partial_\theta + \frac{i}{\sin \theta}\partial_\phi\right)\ .
\end{eqnarray}
By direct evaluation one finds the two torsions to be
\beq
\mathcal{T}_1+i \mathcal{T}_2= \sqrt{2}\pi^*,\qquad
T_1+i T_2= \sqrt{2}\Pi^*
\eeq
where the Newman-Penrose spin coefficient $\pi$ along the curve and its integral $\Pi$ are explicitly
\beq
\fl\quad
\pi=\frac{i}{\sqrt{2}}\frac{a\sin \theta}{(r -i a \cos \theta)^2}\ , \quad
\Pi=\frac{i}{\sqrt{2}}\frac{a(r-r_+)\sin \theta }{(r -i a \cos \theta)(r_+-ia\cos \theta)}\ .
\eeq
The corresponding parallel transported frame is then completely determined.

\section{Concluding remarks}

Using the Newman-Penrose formalism we have constructed a Frenet-Serret frame and corresponding curvature and torsion scalars along a generic null world line, very similar to the case of a timelike world line except that one must generalize the osculating plane to an osculating hyperplane in the null case, as is well known.
For a null Killing trajectory the Frenet-Serret scalars are constant and one may easily find a parallel transported quasi-orthogonal frame along it. Such a frame is also easily constructed for any null geodesic in a general spacetime using any initial Frenet-Serret frame along the geodesic.
The Kerr black hole spacetime provides an explicit example where the frame is easily evaluated explicitly both for accelerated circular orbits and for the geodesic integral curves of the null principal directions.

\section*{Acknowledgments}

We are grateful to Professors Giorgio Ferrarese and Bartolom\'e Coll for useful discussion and suggestions.

\end{document}